\documentclass[aps,floatfix,preprint,preprintnumbers,nofootinbib,superscriptaddress,natbib]{revtex4}
\usepackage{graphicx,float}% Include figure files
\usepackage{epsfig}		
\usepackage{dcolumn}% Align table columns on decimal point
\usepackage{bm}% bold math
\usepackage{subfigure}

\usepackage[all]{xy}
\usepackage{amsmath,upgreek}
\usepackage{amssymb}

\usepackage{pdfpages}
\usepackage{color}
\usepackage{graphicx,epstopdf}
\usepackage[colorlinks=true,
 linkcolor=red,
 urlcolor=purple,
 citecolor=blue,hyperindex]{hyperref}
\def\0{\mbox{\tiny $0$}}
\def\1{\mbox{\tiny $1$}}
\def\2{\mbox{\tiny $2$}}
\def\3{\mbox{\tiny $3$}}
\def\4{\mbox{\tiny $4$}}
\def\5{\mbox{\tiny $5$}}
\def\6{\mbox{\tiny $6$}}
\def\7{\mbox{\tiny $7$}}
\def\8{\mbox{\tiny $8$}}
\def\9{\mbox{\tiny $9$}}

\def\f14{\mbox{\tiny $\frac{1}{4}$}}

\begin{document}

\title{Distorted stability pattern and chaotic features for quantized prey-predator-like dynamics}

\renewcommand{\baselinestretch}{1.2}
\author{A. E. Bernardini}
\email{alexeb@ufscar.br}
\altaffiliation[On leave of absence from]{~Departamento de F\'{\i}sica, Universidade Federal de S\~ao Carlos, PO Box 676, 13565-905, S\~ao Carlos, SP, Brasil.}
\author{O. Bertolami}
\email{orfeu.bertolami@fc.up.pt}
\altaffiliation[Also at~]{Centro de F\'isica do Porto, Rua do Campo Alegre 687, 4169-007, Porto, Portugal.} 
\affiliation{Departamento de F\'isica e Astronomia, Faculdade de Ci\^{e}ncias da
Universidade do Porto, Rua do Campo Alegre 687, 4169-007, Porto, Portugal.}
\date{\today}

\begin{abstract}
Non-equilibrium and instability features of prey-predator-like systems associated to topological quantum domains emerging from a quantum phase-space description are investigated in the framework of the Weyl-Wigner quantum mechanics.
Reporting about the generalized Wigner flow for one-dimensional Hamiltonian systems, $\mathcal{H}(x,\,k)$, constrained by $\partial^2 \mathcal{H} / \partial x \, \partial k = 0$, the prey-predator dynamics driven by Lotka-Volterra (LV) equations is mapped onto the Heisenberg-Weyl non-commutative algebra, $[x,\,k] = i$, where the canonical variables $x$ and $k$ are related to the two-dimensional LV parameters, $y = e^{-x}$ and $z = e^{-k}$. 
From the non-Liouvillian pattern driven by the associated Wigner currents, hyperbolic equilibrium and stability parameters for the prey-predator-like dynamics are then shown to be affected by quantum distortions over the classical background, in correspondence with non-stationarity and non-Liouvillianity properties quantified in terms of Wigner currents and Gaussian ensemble parameters.
As an extension, considering the hypothesis of discretizing the time parameter, non-hyperbolic bifurcation regimes are identified and quantified in terms of $z-y$ anisotropy and Gaussian parameters. The bifurcation diagrams exhibit, for quantum regimes, chaotic patterns highly dependent on Gaussian localization.
Besides exemplifying a broad range of applications of the generalized Wigner information flow framework, our results extend, from the continuous (hyperbolic regime) to discrete (chaotic regime) domains, the procedure for quantifying the influence of quantum fluctuations over equilibrium and stability scenarios of LV driven systems.
\end{abstract}

\date{\today}
\maketitle

\section{Introduction}

The Lotka-Volterra (LV) equations for prey-predator systems \cite{LV1,LV2} were originally proposed as the setup for a classical deterministic description for the ecological equilibrium of competitive populations.
Due to their accessible mathematical properties, species coexistence chains could be straightforwardly described by phase-space closed orbits parameterized by a nonlinear Hamiltonian expressed in terms of the dimensionless one-dimensional $x-k$ phase-space by \cite{Novo2021B}
\begin{equation}\label{Ham}
\mathcal{H}(x,\,k) = a \,x + k + a\, e^{-x} + e^{-k}=\epsilon,
\end{equation}
where $x$ and $k$ variables are correlated with the numbers of prey and predator species, $y$ and $z$, by $y = e^{-x}$ and $z = e^{-k}$, respectively, and $\epsilon$ is an arbitrary constant.
On the phenomenological perspective, while noticing that natural environments generally consist of heterogeneous domains which affect the behavior of microscopic bio-systems, homogeneous prey-predator-like distributions in the domain space are just hypothetical. In Nature, macroscopic and microscopic species are closely linked to their environment conditions, which define growth directives to the categories of populations. Equivalent rate of growth of species can coexist with different strategies related to additional degrees of freedom, even if, for instance, the environment is homogeneous.
Such aspects, once expressed in terms of phenomenological parameters, can turn the stable coexistence of species
into unstable scenarios.

Generically, a stable classical setup as described by the Hamiltonian Eq.~\eqref{Ham} can be modified in order to encompass more complex prey-predator-like configurations which include, for instance, extinction and revival mechanisms \cite{Novo2021B,PRE-LV}, perpetual coexistence \cite{PRE-LV2}, competition-induced chaos \cite{PP00,PP01,PP02} and microscopic molecular dynamics for symbiotic synchronization \cite{PP03,PP04}.

Specifically for microscopic systems, mechanisms that drive both quantum fluctuations and non-linear effects may be admitted, even if a theoretical connection with crude biochemical and biological evolutionary modeling is still unveiled. 
In fact, prey-predator-like oscillations, competition-induced chaos, and symbiotic synchronization are examples of such microscopic behavior which have been identified experimentally \cite{Nature01,PP00,PP02,PP03,PP04}. They can be regarded as a motivation for considering the quantization of LV systems and possibly to respond how and at which scales classical macroscopic and quantum microscopic evolution coexist, and if quantum effects arise at measurable scales.
The answer for these questions can also be relevant, for instance, in the investigation of the above quoted stability criteria for microbiological communities \cite{Nature01, Nature02}, or even in the analysis of stochastic system dynamics \cite{Allen,Grasman,PRE-LV4} and in the description of phase transitions in finite microscopic systems \cite {PRE-LV3}.
Therefore, despite the classical background driven by LV equations, the inclusion of both quantum-like and instability triggers must be considered, in particular, in the scope of equilibrium and stability analysis

For macroscopic (non-quantized) ecosystems \cite{PRE-LV,SciRep02,PRE-LV2,RPSA-LV}, a quantitative description, either in terms of the hyperbolic equilibrium description \cite{HG,Book,Book2}, for a presumed continuous variable dynamics, or in terms of chaos pattern classification, by the Hopf bifurcation analysis (if the predator-prey dynamics is not covered by the hyperbolic equilibrium regime) is already admitted.
Otherwise, for microscopic organisms or bio-systems, the understanding of the transition between classical and quantum regimes demands for the inclusion of systematic and operative procedures which are still incipient \cite{Novo2021,Novo2021B,Steuernagel3}.

In the framework here considered, non-equilibrium and instability properties are associated to topological quantum domains which emerge from a quantum phase-space description of the prey-predator-like dynamics \cite{Novo2021}.
Generically, for one-dimensional Hamiltonian systems (cf. Eq.~(\ref{Ham})), quantum features associated to phase-space patterns, stationary behavior, and information fluxes are straightforwardly quantified in terms of Wigner currents and related properties \cite{Steuernagel3,Wigner,Ballentine,Case,Meu2018,Zurek02,NossoPaper}.

The classical dynamics is given by the equations of motion obtained from Hamiltonian Eq.~(\ref{Ham}), i.e.
\begin{eqnarray}\label{Ham2}
d{x}/d\tau &=& \{x,\mathcal{H}\}_{PB} = 1-e^{-k},\\
d{k}/d\tau &=& \{k,\mathcal{H}\}_{PB} = e^{-x} - 1,
\end{eqnarray}
where $\tau$ is the dimensionless time, and the prey-predator $z - y$ system mapped into $z \mapsto e^{-k}$ and $y \mapsto e^{-x}$ can be exactly evaluated in terms of phase-space coordinates through the Weyl-Wigner (WW) framework \cite{Novo2021B}. This procedure allows for an effective quantum description for the system \cite{Novo2021}.
Equilibrium and stability conditions affected by quantum distortions over the classical background can then be theoretically connected to stationarity and Liouvillianity drivers \cite{Steuernagel3,Wigner,Ballentine,Case,Meu2018,Zurek02,NossoPaper}. These quantum features can then be quantified in terms of Wigner currents and ensemble parameters.

Conceptually, the collective behavior depicted from phase-space effective quantum trajectories are interpreted as averaged-out results of the space-time evolution of the species distributions.
Quantum deviations from classical patterns and their effects on the time evolution of the prey-predator number of species can then be evaluated.
In fact, the WW framework allows for identifying how classical and quantum evolution coexists at different scales and how prey-predator {\em quantum analogue} effects emerge \cite{Novo2021B,Novo2021C}.

Considering that the prey-predator equilibrium regime is driven by an autonomous system of ordinary differential equations, our analysis here is focused on verifying if the conditions for equilibrium and stability criteria are met.
As it will be verified, the quantum distortions over the classical pattern, once convoluted by Gaussian ensembles \cite{Novo2021B}, produce evident hyperbolic and non-hyperbolic equilibrium and instability patterns which can all be quantified with tools of the WW framework.

Having set the goals of our work, the outline of the manuscript is as follows.
Sec.~II is concerned with the foundations of the generalized WW framework \cite{Novo2021}, which result in the quantum driven phase-space trajectories given in terms of Wigner currents.
Stationarity and Liouvillianity quantifying operators obtained in the context of the extended Wigner framework for non-linear Hamiltonians, $\mathcal{H}(x,\,k) = \mathcal{K}(x) + \mathcal{V}(k)$, are recovered \cite{Novo2021} and the mathematical structure for obtaining the corresponding statistically convoluted Gaussian ensemble exact solutions is re-examined.
Non-equilibrium and instability features of prey-predator-like competition systems associated to topological quantum domains emerging from such a quantum phase-space description are investigated in Sec.~III.
Since the equilibrium regime is generated by an autonomous system of ordinary differential equations from the WW phase-space, the quantitative correspondence of stationarity and Liouvillianity with stability properties in terms of hyperbolic equilibrium parameters is discussed. Equilibrium and stability quantifiers can thus be obtained for quantum Gaussian ensembles when they are dynamically driven by the corresponding Winger currents.
Furthermore, emergent topological phases due to the diffusive appearance of unstable vortices and saddle-points in the phase-space, as an effect of quantum distortions over the classical background, are also identified.
Finally, considering the hypothesis of discretizing the time parameter, beyond the hyperbolic regime, non-hyperbolic bifurcation patterns are identified in Sec.~IV, being described in terms of species anisotropy and Gaussian localization parameters. 
Our conclusions are drawn in Sec.~V, where the bridges between classical and quantum descriptions of prey-predator-like systems are evinced, and a consistent interpretation on the meaning of quantum distortions is provided.

\section{Quantum driven phase-space trajectories}

The WW phase-space framework \cite{Wigner,Ballentine,Case} encompasses all the quantum features (see the Appendix I) through a {\em quasi}-probability distribution function of canonical coordinates of position, $x$, and momentum, $k$, through the so-called Wigner function, $\mathcal{W}(x,\, k)$. Since it is associated with the quantum density matrix operator, $\hat{\rho} = |\psi \rangle \langle \psi |$, through its Weyl transform one has
\begin{equation}
\hat{\rho} \to \mathcal{W}(x,\, k) = \pi^{-1} 
\int^{+\infty}_{-\infty} \hspace{-.35cm}dw\,\exp{\left[2\, i \, k \,w\right]}\,
\psi(x - w)\,\psi^{\ast}(x + w),\label{222}
\end{equation}
written in a dimensionless form, where the Planck constant, $\hbar$, has been set equal to $1$.
Correspondently, the probability flux evolves according to the continuity equation written as \cite{Case,Ballentine,Steuernagel3,NossoPaper,Meu2018}
\begin{equation}\label{z51dim}
{\partial_{\tau} \mathcal{W}} + {\partial_x \mathcal{J}_x}+{\partial_k \mathcal{J}_k} = {\partial_{\tau} \mathcal{W}} + \mbox{\boldmath $\nabla$}_{\xi}\cdot\mbox{\boldmath $\mathcal{J}$} =0,
\end{equation}
where the time, $\tau$, is also dimensionless. For generic Hamiltonians in the form of $\mathcal{H}(x,\,k) = \mathcal{K}(k) + \mathcal{V}(x)$, 
the associated Wigner currents are straightforwardly given by \cite{Novo2021}
\begin{eqnarray}
\label{imWAmm}\mathcal{J}_x(x, \, k;\,\tau) &=& +\sum_{\eta=0}^{\infty} \left(\frac{i}{2}\right)^{2\eta}\frac{1}{(2\eta+1)!} \, \left[\partial_k^{2\eta+1}\mathcal{K}(k)\right]\,\partial_x^{2\eta}\mathcal{W}(x, \, k;\,\tau),\\
\label{imWBmm}\mathcal{J}_k(x, \, k;\,\tau) &=& -\sum_{\eta=0}^{\infty} \left(\frac{i}{2}\right)^{2\eta}\frac{1}{(2\eta+1)!} \, \left[\partial_x^{2\eta+1}\mathcal{V}(x)\right]\,\partial_k^{2\eta}\mathcal{W}(x, \, k;\,\tau),
\end{eqnarray}
so to account for the quantum back reaction through the $\eta > 1$ contributions from the corresponding series expansions.
Once truncated at $\eta = 0$, currents Eqs.~\eqref{imWAmm} and \eqref{imWBmm} yield the classical Liouvillian regime \cite{Case,Ballentine}.

For phase-space coordinates identified by $\mbox{\boldmath $\xi$} = \xi_x \hat{x} + \xi_k \hat{k}$, stationarity and Liouvillianity can be locally quantified by divergence operators given by, $\mbox{\boldmath $\nabla$}_{\xi}\cdot\mbox{\boldmath $\mathcal{J}$}$ and $\mbox{\boldmath $\nabla$}_{\xi} \cdot \mathbf{w}$, respectively \cite{Steuernagel3,NossoPaper,Meu2018}. These are mutually connected by a parametric definition of a quantum-analog velocity, $\mathbf{w}$, implicitly given in terms of the vector currents, $\mbox{\boldmath $\mathcal{J}$} = \mathbf{w}\,\mathcal{W}$\footnote{From which it can be noticed that $\mbox{\boldmath $\nabla$}_{\xi}\cdot\mbox{\boldmath $\mathcal{J}$} = \mathcal{W}\,\mbox{\boldmath $\nabla$}_{\xi}\cdot\mathbf{w}+ \mathbf{w}\cdot \mbox{\boldmath $\nabla$}_{\xi}\mathcal{W}$.},which has the classical limit, $\mathbf{w} \to \mathbf{v}_{\xi(\mathcal{C})}$,consistently described by the vector velocity $\mathbf{v}_{\xi(\mathcal{C})} = \dot{\mbox{\boldmath $\xi$}} = (\dot{x},\,\dot{k})\equiv ({\partial_k \mathcal{H}},\,-{\partial_x \mathcal{H}})$. From those definitions, the associated stationarity and Liouvillianity divergence operators are given respectively by
\begin{equation} \label{helps}
\mbox{\boldmath $\nabla$}_{\xi}\cdot\mbox{\boldmath $\mathcal{J}$} = \sum_{\eta=0}^{\infty}\frac{(-1)^{\eta}}{2^{2\eta}(2\eta+1)!} \, \left\{
\left[\partial_x^{2\eta+1}\mathcal{V}(x)\right]\,\partial_k^{2\eta+1}\mathcal{W}
-
\left[\partial_k^{2\eta+1}\mathcal{K}(k)\right]\,\partial_x^{2\eta+1}\mathcal{W}
\right\},\end{equation}
and
\begin{equation}\label{div2}
\mbox{\boldmath $\nabla$}_{\xi} \cdot \mathbf{w} = \sum_{\eta=0}^{\infty}\frac{(-1)^{\eta}}{2^{2\eta}(2\eta+1)!}\left\{
\left[\partial_k^{2\eta+1}\mathcal{K}(k)\right]\,
\partial_x\left[\frac{1}{\mathcal{W}}\partial_x^{2\eta}\mathcal{W}\right]
-
\left[\partial_x^{2\eta+1}\mathcal{V}(x)\right]\,
\partial_k\left[\frac{1}{\mathcal{W}}\partial_k^{2\eta}\mathcal{W}\right]
\right\}, ~~~\end{equation}
from which, stationary and classical patterns are recovered by $\mbox{\boldmath $\nabla$}_{\xi}\cdot\mbox{\boldmath $\mathcal{J}$}=0$ and $\mbox{\boldmath $\nabla$}_{\xi} \cdot \mathbf{w}=0$, respectively \cite{Meu2018}.

From this point, replacing the Wigner distribution, $\mathcal{W}(x, \, k;\,\tau)$, by a Gaussian one\footnote{For the classical to quantum transition viewed by the statistical point of view, the classical analog of the second-order moments of position and momentum coordinates, which define the Heisenberg uncertainty principle, should consistently satisfy 
the same constraints of position and momentum quantum observables. This procedure is isomorphic to the implementation of Gaussian ensembles (or states) \cite{Ballentine,PRAPRAPRA} in the sense that, in both cases, the same statistics is reproduced.}, 
\begin{equation}
\mathcal{G}_{\alpha}(x(\tau), \, k(\tau)) \equiv \mathcal{G}_\alpha(x,\,k) = \frac{\alpha^2}{\pi}\, \exp\left[-\alpha^2\left(x^2+ k^2\right)\right],
\end{equation}
into Eqs.~\eqref{imWAmm} and \eqref{imWBmm}, allows for writing
\cite{Novo2021}
\begin{equation}\label{ssae}
\partial_\chi^{2\eta+1}\mathcal{G}_{\alpha}(x, \, k) = (-1)^{2\eta+1}\alpha^{2\eta+1}\,\mbox{\sc{h}}_{2\eta+1} (\alpha \chi)\, \mathcal{G}_{\alpha}(x, \, k),\qquad\mbox{for $\chi = x,\, k$,}
\end{equation}
where $\mbox{\sc{h}}_n(\alpha \chi)$ are the Hermite polynomials of order $n$.

Also considering the auxiliary derivatives for $\mathcal{K}(k)$ and $\mathcal{V}(x)$ obtained from Eq.~\eqref{Ham} as
\begin{eqnarray}
\label{t111B}
\partial_x^{2\eta+1}\mathcal{K}(k) &=& \delta_{\eta 0} - e^{-k},\\
\label{t222B}
\partial_k^{2\eta+1}\mathcal{V}(x) &=& a \left(\delta_{\eta 0} - e^{-x}\right), 
\end{eqnarray}
the Gaussian convoluted prey-predator dynamics can thus be described in terms of Wigner currents written in the form of \cite{Novo2021B},
\begin{eqnarray}
\label{imWA4CC3mm}
\partial_x\mathcal{J}^{\alpha}_x(x, \, k) &=& -2 \left[\alpha^2 \,x - 
\sin\left(\alpha^2\,x\right)\,e^{\frac{\alpha^2}{4}-k}
\right]
\mathcal{G}_{\alpha}(x, \, k),\\
\label{imWB4CC3mm}
\partial_k\mathcal{J}^{\alpha}_k(x, \, k) &=& +2a\left[\alpha^2\,k - 
\sin\left(\alpha^2\,k\right)\,e^{\frac{\alpha^2}{4}-x}
\right]
\mathcal{G}_{\alpha}(x, \, k),
\end{eqnarray}
which correspond to the convergent expression for the series expansions, Eqs.~\eqref{imWAmm} and \eqref{imWBmm}, since the convergent contributions
\begin{equation}
\sum_{\eta=0}^{\infty}\mbox{\sc{h}}_{2\eta+1} (\alpha \chi)\frac{s^{2\eta+1}}{(2\eta+1)!} = \sinh(2s\,\alpha\chi) \exp[-s^2]
\end{equation}
have been found \cite{Novo2021,Novo2021B}.
Finally, from the above results, and after a straightforward integration, the quantum analog Wigner velocity components are given by
\begin{eqnarray}
\label{imWA4CCD4mm}\mbox{w}^{\alpha}_x(x, \, k) &=& 
1
-\frac{i\,\sqrt{\pi}}{2\alpha} \,e^{-(k-\alpha^2 x^2)}
\left\{\mbox{\sc{Erf}}\left[\alpha(x-i/2)\right]-\mbox{\sc{Erf}}\left[\alpha(x+i/2)\right]\right\},\\
\label{imWB4CCD4mm}\mbox{w}^{\alpha}_k(x, \, k) &=& 
-a\,\left\{1-
\frac{i\,\sqrt{\pi}}{2\alpha} \,e^{-(x-\alpha^2 k^2)}
\left\{\mbox{\sc{Erf}}\left[\alpha(k-i/2)\right]-\mbox{\sc{Erf}}\left[\alpha(k+i/2)\right]\right\}\right\},\,\,\,\,
\end{eqnarray}
written in terms of Gaussian error functions, $\mbox{\sc{Erf}}[\dots]$.

The above results for the Wigner flow pattern can all be constrained by the form of $\mbox{\boldmath $\mathcal{J}^{\alpha}$} = \mathbf{w}\,\mathcal{G}_{\alpha}(x, \, k)$, with $\mathbf{w}=(\mbox{w}^{\alpha}_x,\,\mbox{w}^{\alpha}_k)$, which, besides yielding analytical results for components of the Wigner current, $\mbox{\boldmath $\mathcal{J}$}$, bring up the topological properties of the Wigner flow and their consequences to the prey-predator-like dynamics, as will be discussed in the following.

\subsection{Results for Gaussian ensembles}

\begin{figure}
\vspace{-1.2 cm}\includegraphics[scale=0.10]{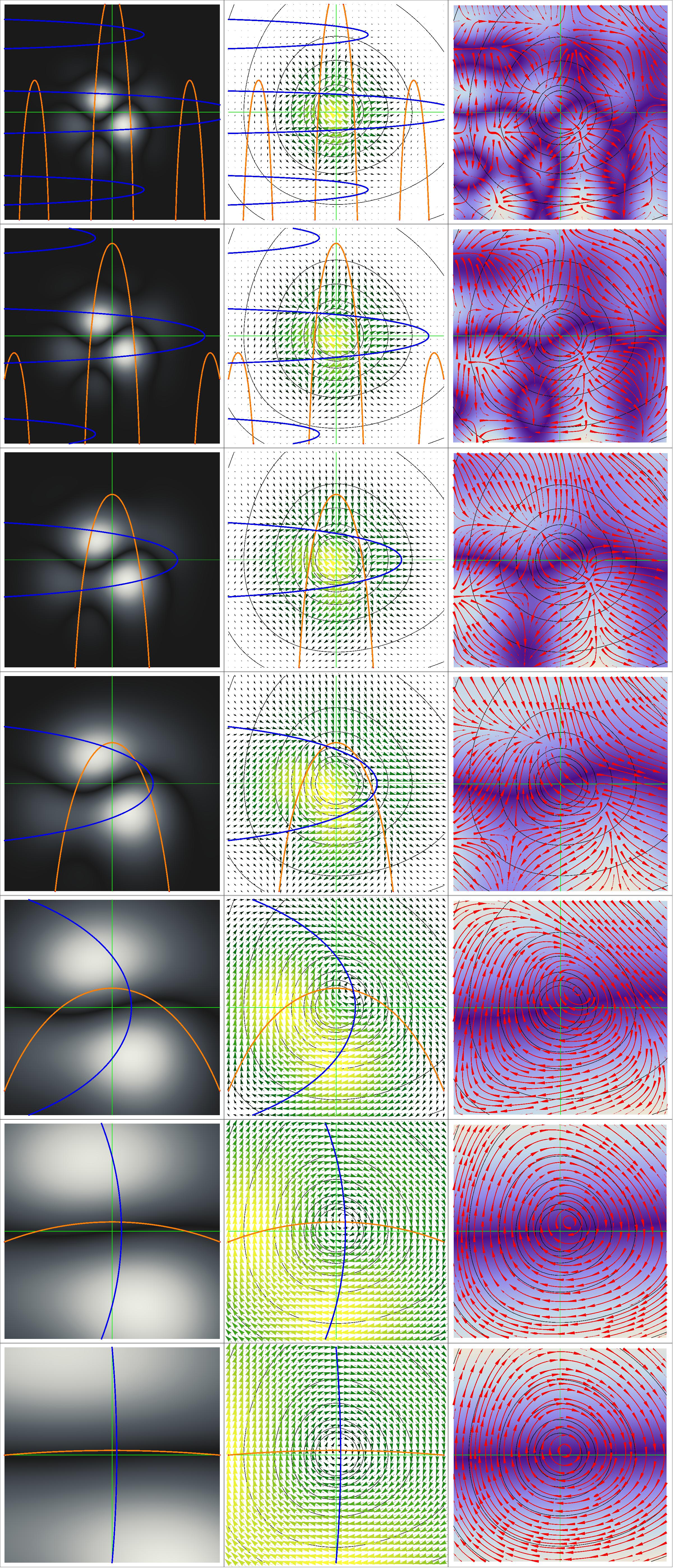}
\renewcommand{\baselinestretch}{.6}
\vspace{-.2 cm}\caption{\footnotesize
(Color online)
{\em First column}: Stationarity quantifier, $\mbox{\boldmath $\nabla$}_{\xi} \cdot \mbox{\boldmath $\mathcal{J}$}$ for the Gaussian ensemble. The color scheme depicts {\em quasi}-stable (darker regions) and unstable (lighter) regions. Increasing values of $\alpha$ localizes the quantum distortions, which result into evinced non-stationarity (unstable) domains. Blue and orange contour lines are for $\mathcal{J}_x =0$ and $\mathcal{J}_k =0$, respectively.
{\em Second column}: Corresponding Wigner flux, $\mbox{\boldmath $\mathcal{J}$}$. The vector plot scheme is modulated by $\vert\mbox{\boldmath $\mathcal{J}$}\vert^{1/4}$ and the {\em greenyellow} color scheme depicts the intensity of the Wigner current from maximal (yellow) to minimum (green).
The quantum critical points are identified by orange-blue crossing lines, for $\mathcal{J}^{(2)}_x = \mathcal{J}^{(2)}_k=0$.
{\em Third column}: Liouvillianity quantifier, $\mbox{\boldmath $\nabla$}_{\xi} \cdot \mathbf{w}$, superposed by the normalized quantum velocity field (red arrows), $\mathbf{w} /\vert\mathbf{w}\vert$.
The divergence values, $\mbox{\boldmath $\nabla$}_{\xi} \cdot \mathbf{w}$, are depicted through the background color scheme, from darker regions, $\mbox{\boldmath $\nabla$}_{\xi} \cdot \mathbf{w} \sim 0$, to lighter regions, $\mbox{\boldmath $\nabla$}_{\xi} \cdot \mathbf{w} > 0$.
All the results are for $\alpha^2 = 10$ (first row), $8,\,6,\,4,\,2,\,1$ and $1/2$ (last row) and for the $x$ (horizontal axis) and $k$ (vertical axis) isotropic case ($a=1$), with $k,\,x\in[-3/2,+3/2]$ (unnecessary print values are suppressed from the axis as to have clearer visual effect).
Classical trajectories are shown as a collection of background black lines.}
\label{Bio01}
\end{figure}

The phase-space prey-predator Gaussian ensemble Wigner flow pattern evaluated in terms of the Gaussian parameter, $\alpha$, from $\alpha^2 = 1/2$ to $10$ is depicted in Fig.~\ref{Bio01}. 
The plots are for the stationarity quantifier (first column), $\mbox{\boldmath $\nabla$}_{\xi} \cdot \mbox{\boldmath $\mathcal{J}$}$, for the Wigner flux (second column), $\mbox{\boldmath $\mathcal{J}$}$, and for the Liouvillian quantifier (third column), $\mbox{\boldmath $\nabla$}_{\xi} \cdot \mathbf{w}$, which is superposed by the normalized quantum velocity field (red arrows), $\mathbf{w} /\vert\mathbf{w}\vert$.
In a kind of diffusive behavior, discretized quantum domains emerge for (highly) peaked Gaussian envelops, which in Fig.~\ref{Bio01} is evaluated from $\alpha^2 = 10 \gg 1$ (first row) to $\alpha^2 = 1/2 \gtrsim0$ (last row). 
Quantum features are indeed suppressed for smoothly peaked Gaussian envelops. 
For $\alpha^2 \lesssim 1$, the {\em quasi}-stationary quantum pattern approaches the classical regime, and the quantum effects are seen as a small displacement of the equilibrium coordinates, $(x,\,k) \sim (0,0)$, for {\em quasi}-closed orbits. 
The {\em greenyellow} color scheme qualitatively depicts the intensity of the Wigner current from maximal (yellow) to minimum (green) values. In fact, it shows that the probability flux pushes the equilibrium points away from $(x,\,k) \sim (0,0)$ as $\alpha$ increases.
In both first and second columns, additional green and orange contour lines are respectively evinced for $\mathcal{J}_x =0$ and $\mathcal{J}_k =0$, as to depict the boundaries for the reversal of the Wigner flow in the $x$ and $k$ direction, respectively.
For increasing values of $\alpha$, which correspond to more peaked Gaussian distributions (from lower to upper rows), the quantum distortions emerge as (anti)clockwise vortices (with winding number equals to $(-)+1$), as well as separatrix intersections and saddle points (with winding number equals to $0$), i.e. flow stagnation points which are all identified by orange-blue crossing lines, where $\mathcal{J}^{(2)}_x = \mathcal{J}^{(2)}_k=0$.
The contra-flux fringes (bounded by blue/orange lines) destroy the closed orbit patterns since they emerge to compensate the retarded evolution of the quantum flux, and introduce discretized quantum domains. These can be identified by the Liouvillian quantifier (third column), which clearly depicts Wigner quantum distorted velocities (red arrows), $\mathbf{w}$. The Wigner function and current patterns define a kind of phase-transition driven by $\alpha$ as the definitive imprint of the classical to quantum transition. The phase-transition is identified by the emergence of multiple stagnation points, with $\mathcal{J}^{(2)}_x = \mathcal{J}^{(2)}_k=0$, for multiple quantum domains connected by the topological features of the Wigner flow.

\section{Hyperbolic stability} 

\begin{figure}
\vspace{-1.2 cm}
$a)$\includegraphics[scale=0.45]{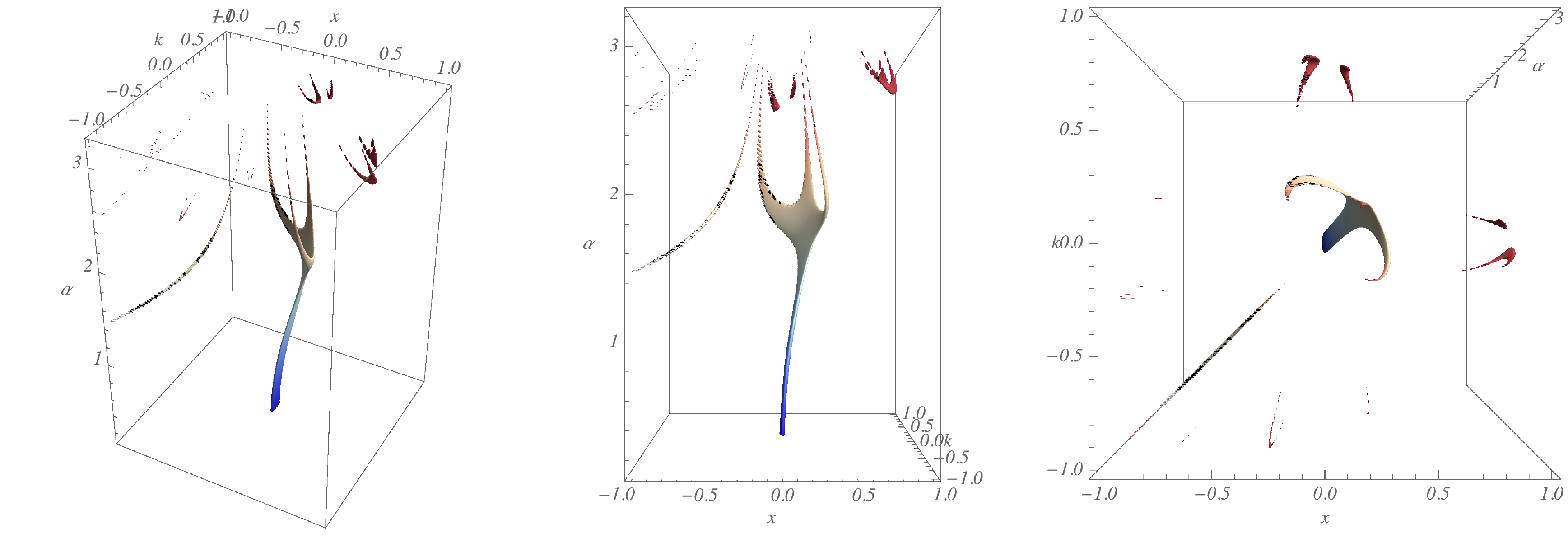}
$b)$\includegraphics[scale=0.45]{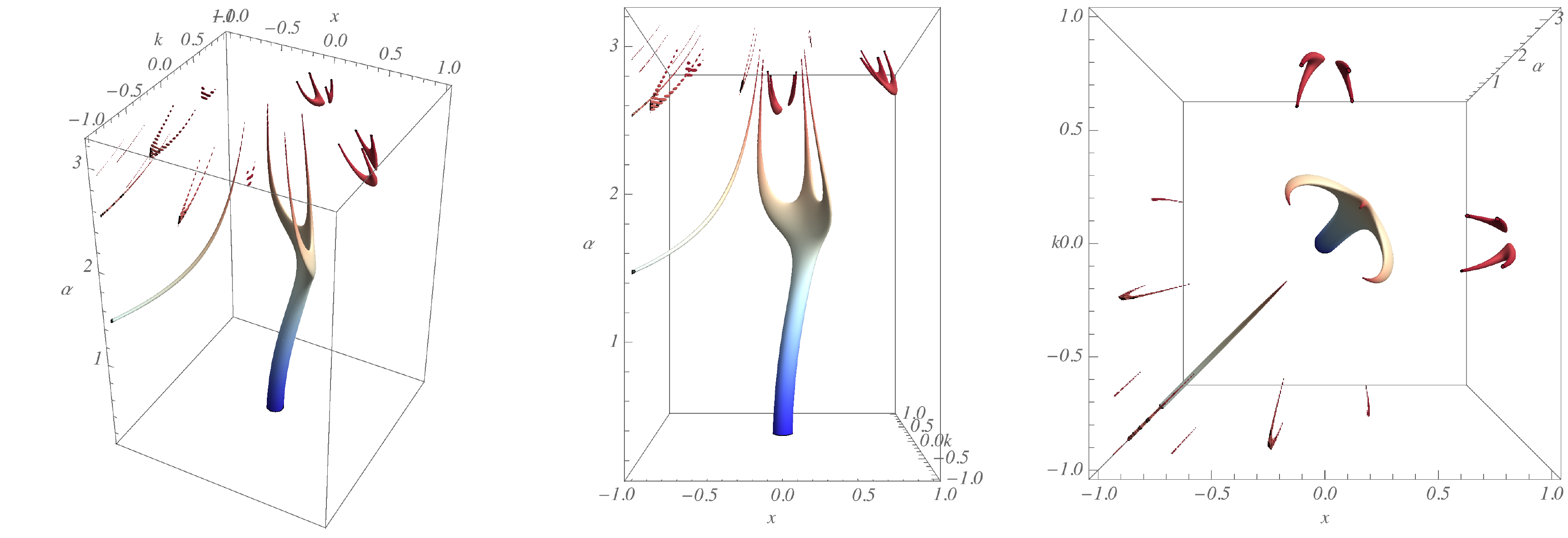}
$c)$\includegraphics[scale=0.45]{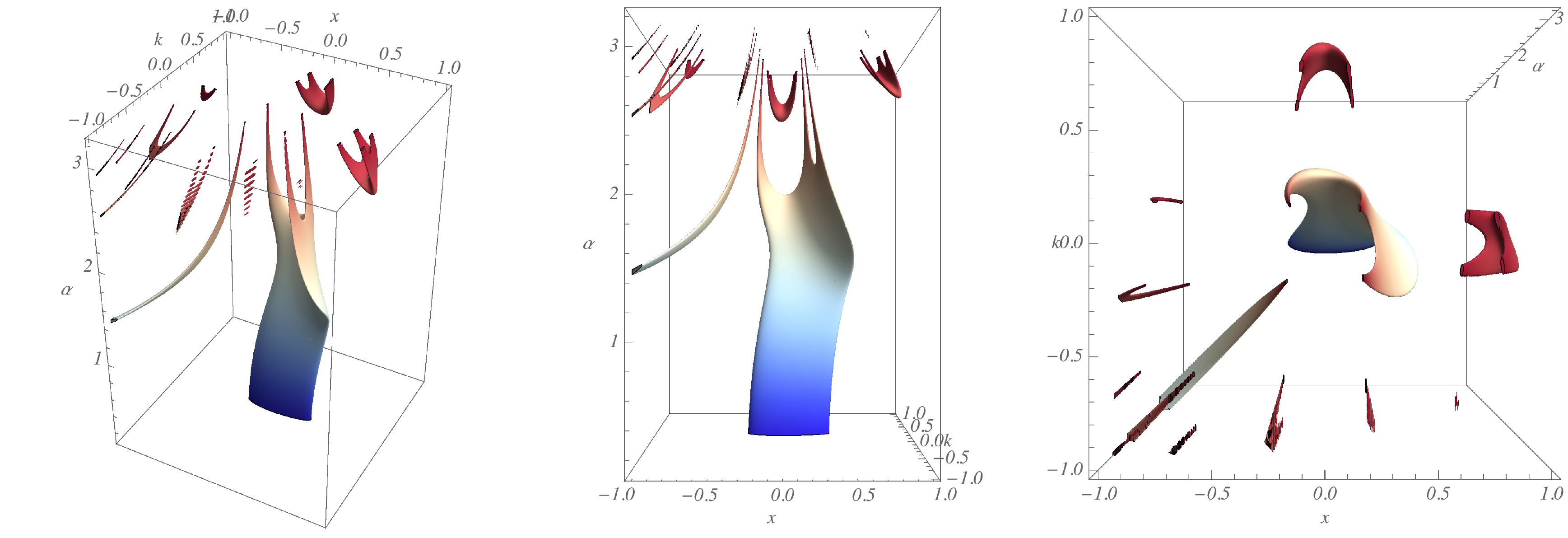}
\renewcommand{\baselinestretch}{.6}
\caption{\footnotesize
(Color online)
Region plot scheme for the phase-space evolution of stagnation points in the phase-space, $x-k$.
The plots show how the attractors (blue regions) are affected by the magnitude of the Gaussian spreading parameter $\alpha$, 
for $0 \lesssim \alpha \lesssim 3.2$. The equilibrium point (flux) surrounding envelop is defined by $\vert\mathbf{w}\vert < 0.006$, and the plots are exhibited for: $a)$ $a = 4$ (first row), $b)$ $a = 1$ (second row) and $c)$ $a=1/4$ (third row). 
for $0 < \alpha \lesssim 1.825$, the anisotropy parameter, $a$, drives the stability regime for attractor focus and node points. For $a > 1$ one shall have stable regimes, as it appears for the convergent behavior of the equilibrium points for $a = 4$ (first row).
Conversely, for $a < 1$ one shall unstable regimes, as it appears for the diffusive behavior of the equilibrium points for $a = 1/4$ (third row).
Results are for the Wigner flow in correspondence with Fig.~\ref{Bio01}. Local effects compensate each other when sliced views of the Wigner flux for fixed $\alpha$ are considered, i.e. either when two vortices of opposite winding numbers match each other or when saddle points (white to red regions) mutually annihilates each other.
The portraits are the same for different angle views.
\label{Bio0002}
}
\end{figure}

\hspace{1em} The stagnation points which are identified in the phase-space Wigner flow (cf. Fig.~\ref{Bio01}) can be evaluated in terms of the Gaussian parameter, $\alpha$, which implicitly accounts for the contributions due to the quantum fluctuations over the classical background, according to the magnitude of the Gaussian localization, which, for $\alpha = 0$, reproduces the classical limit of Eqs.~\eqref{imWA4CCD4mm} and \eqref{imWB4CCD4mm}.
The effects depicted in Fig.~\ref{Bio01}, which emerge from increasing $\alpha$ values, can be identified through the phase-space evolution of additional quantum mechanically emerging equilibrium points which replace the unique classical equilibrium point, $(x,\,k) \sim (0,0)$. 
The rows of in Fig.~\ref{Bio0002} depict three different frame views for the behavior of the phase-space stagnation points in terms of $\alpha$, for $0 \lesssim \alpha \lesssim 3.2$. 

The equilibrium point (flux) surrounding envelop is defined by $\vert \mathbf{w}\vert < 0.006$, and the plots are exhibited for $a = 4$ (first row), $a = 1$ (second row) and $a=1/4$ (third row). 
As it will be explained in the following, according to the so-called hyperbolic equilibrium criteria, for $0 < \alpha \lesssim 1.825$, the anisotropy parameter, $a$, drives the stability regime for the attractor focus and node points. For $a > 1$, one has stable regimes, as it appears for the convergent behavior of the equilibrium points for $a = 4$ (first row).
Conversely, for $a < 1$, one has unstable regimes, as it appears for the diffusive behavior of the equilibrium points for $a = 1/4$ (third row).
The spreading behavior of the Gaussian ensemble, which runs from red bubble saddle-point islands to the blue enveloped focus, corresponding to decreasing values of $\alpha$, recovers a {\em quasi} classical pattern, for which the quantum imprint is found only for the small displacement (from $(x,\,k) \sim (0,0)$) of the attractor equilibrium points. The blue region corresponds to the hyperbolic equilibrium regime, and can be described, for instance, through a perturbative analysis \cite{Novo2021C}. 

In fact, for clarifying the above statements, one should turn attention to the definition of the equilibrium points of a two-dimensional dynamical system. In the context of the classical phase-space dynamics, they correspond to the solutions of a system of ordinary differential equations with stationary behavior. 
Therefore, the equilibrium is geometrically defined by $\dot{\mbox{\boldmath{$\xi$}}} = 0$ (i.e. $\mbox{v}_{x(\mathcal{C})}=\mbox{v}_{k(\mathcal{C})}=0$), whichhas a straightforward quantum correspondence expressed in terms of the quantum velocity, $\mathbf{w}$, by $\mbox{w}_x=\mbox{w}_k=0$. The equilibrium point solutions, in both classical and quantum descriptions, correspond to the phase-space stagnation points. For a continuous system reduced to an equivalent discretized iterative system, they correspond to the so-called fixed points of the equation system.
Considering the effective quantum dynamics described by the behavior of the phase-space velocities as
\begin{eqnarray}
\mbox{w}_x &=& \mathcal{J}_x(x,\,k;\,t)/\mathcal{W}(x,\,k,\,;t) \equiv f(x,\,k),\nonumber\\
\mbox{w}_k &=& \mathcal{J}_x(x,\,k;\,t)/\mathcal{W}(x,\,k,\,;t) \equiv g(x,\,k),
\end{eqnarray}
the so-called hyperbolic stability of the equilibrium points is established by the features of the Jacobian matrix identified by
\begin{equation}
j (x,\,k) = \left[\begin{array}{rr}
\partial_x f(x,\,k) & \partial_k f(x,\,k)\\
\partial_x g(x,\,k) & \partial_k g(x,\,k)
\end{array}\right],
\end{equation}
which defines an approximated linear stability by means of its eigenvalues so that equilibrium and stability conditions can be naturally stratified into subclassifications, through the trace, ${Tr}[\dots]$, and the determinant, ${Det}[\dots]$, of $j (x,\,k)$, when all derivatives are evaluated at the equilibrium point, ${\mbox{\boldmath{$\xi$}}_e}= (x_e,\,k_e)$, obtained from $f(x_e,\,k_e)=g(x_e,\,k_e)=0$.

To summarize, $j(x_e,\,k_e)$ has all the eigenvalues with negative real parts for asymptotically stable systems. 
For unstable systems, at least one eigenvalue of $j(x_e,\,k_e)$ has a positive real part. 
Likewise, the Jacobian matrix establishes the conditions for the so-called hyperbolic equilibrium if all their corresponding eigenvalues have non-zero real parts. 
Finally, if at least one eigenvalue of the Jacobian matrix at equilibrium points, $j(x_e,\,k_e)$, has a zero real part, then the equilibrium is not hyperbolic, and the robustness of equilibrium and stability conditions require another criterion \cite{Book}.
Topologically, through the observation of the vector field distribution, $\mathbf{w}(x,\,k)$, at${\mbox{\boldmath{$\xi$}}_e}= (x_e,\,k_e)$, the hyperbolic equilibrium and stability features are as follows: saddle points, which correspond to unstable configurations, for real eigenvalues with opposite signs; divergent (unstable) nodes, for both real eingenvalues with positive signs; convergent (stable) nodes, for both real eingenvalues with negative signs; divergent (unstable) focus, for both complex eingenvalues with positive real parts; and convergent (stable) focus, for both complex eingenvalues with negative real parts.

According to the above criteria, one notices that focus and node stabilities are defined by trace properties as 
\begin{eqnarray}
Tr[j(x_e,\,k_e)] &=& \mbox{\boldmath $\nabla$}_{\xi}\cdot\mathbf{w}\vert_{{\mbox{\boldmath{$\xi$}}_e}} > 0 \qquad \to \mbox{instability},\nonumber\\
Tr[j(x_e,\,k_e)] &=& \mbox{\boldmath $\nabla$}_{\xi}\cdot\mathbf{w}\vert_{{\mbox{\boldmath{$\xi$}}_e}} < 0 \qquad \to \mbox{stability},
\end{eqnarray}
when 
\begin{eqnarray}
Det[j(x_e,\,k_e)] &=& \partial_x f\vert_{{\mbox{\boldmath{$\xi$}}_e}}\,\partial_k g\vert_{{\mbox{\boldmath{$\xi$}}_e}}- \partial_k f\vert_{{\mbox{\boldmath{$\xi$}}_e}}\,\partial_x g\vert_{{\mbox{\boldmath{$\xi$}}_e}} > 0 \qquad \to \mbox{for focus and nodes},
\end{eqnarray}
and
\begin{eqnarray}
Det[j(x_e,\,k_e)] &=&\partial_x f\vert_{{\mbox{\boldmath{$\xi$}}_e}}\,\partial_k g\vert_{{\mbox{\boldmath{$\xi$}}_e}}- \partial_k f\vert_{{\mbox{\boldmath{$\xi$}}_e}}\,\partial_x g\vert_{{\mbox{\boldmath{$\xi$}}_e}}< 0 \qquad \to \mbox{for saddle points}.
\end{eqnarray}
Focus and nodes are also separated by the threshold value, $\Delta[j] = Tr[j]^2 - 4 Det[j] = 0$, with
\begin{eqnarray}
\Delta[j(x_e,\,k_e)] && > 0 \qquad \mbox{for nodes},\nonumber\\
\Delta[j(x_e,\,k_e)] && < 0 \qquad \mbox{for focus}.
\end{eqnarray}

Since the hyperbolic equilibrium admits small linear perturbations over the system of equations, the phase-space portrait qualitatively does not deviate from the steady state configuration.
Hence, the local phase portrait of a nonlinear system can be perturbatively mapped by its linearized version, which equivalently accounts for eventual short displacements of the fixed points (cf. theHartman-Grobman theorem \cite{HG}).
Conversely, several types of non-hyperbolic equilibrium patterns result into local bifurcations, which may change stability, suppress the fixed point features, or even split them into several equilibrium points, as it is qualitatively depicted in Fig.~\ref{Bio0002}.

\begin{figure}
\includegraphics[scale=0.7]{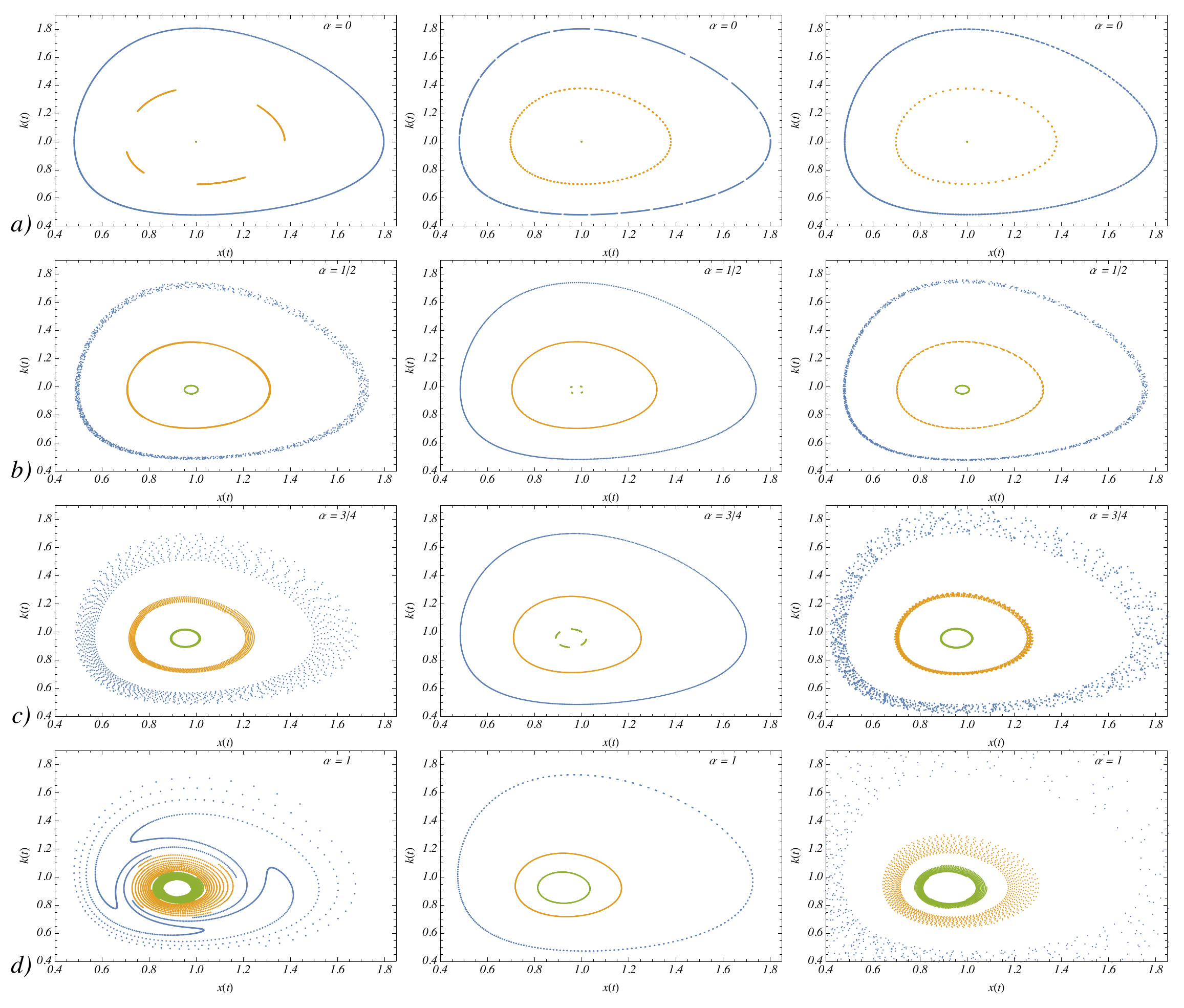}
\renewcommand{\baselinestretch}{.6}
\caption{\footnotesize
(Color online)
Poincar\'e maps for: $a)$ $\alpha =0$ (first row); $b)$ $\alpha =1/2$ (second row); $c)$ $\alpha =3/4$ (third row); $d)$ $\alpha =1$ (forth row), and for $a=1.01$ (first column), $a=1$ (second column) and $a=0.99$ (third column).
\label{Poinc}
}
\end{figure}

Turning back to the results from Fig.~\ref{Bio0002}, which correspond to Eqs.~(\ref{imWA4CCD4mm}) and (\ref{imWB4CCD4mm}), the equilibrium point classifications are covered by the hyperbolic equilibrium criterium only for $\alpha \lesssim 1.825$, where short displacements from classical configurations ($\alpha\sim0$) are evinced by the blue region. 
Despite approaching classical-like closed orbits, they are perturbed by a quantum vortex distortion which emerges from surrounding values of $x$ and $k$ and destroys the (classical) closed orbit pattern.
These features can be noticed from the Poincar\'e maps for time intervals set as a quarter the orbit period, $T/4$, as depicted in Fig.~\ref{Poinc} for $\alpha =0$ (first row), $\alpha =1/2$ (second row), and $\alpha =3/4$ (third row), all under the hyperbolic regime.
The Poincar\'e maps of any continuous dynamical system show successive points of the phase-space trajectory computed by means of the integration of the original continuous system.
From Fig.~\ref{Poinc}, it can be noticed that the relevant deviations from the closed orbit regime emerge due to tiny anisotropic perturbations, which were introduced by setting $a=1.01$ (first column) and $a=0.99$ (third column).
For the isotropic configuration, with $a = 1$, since $Tr[j(x_e,\,k_e)] = 0$ (cf. Eqs.~(\ref{imWA4CCD4mm}) and (\ref{imWB4CCD4mm})), classical ($\alpha = 0$) and quantum ($\alpha\neq 0$) closed orbits are not affected.
Correspondently, stable and unstable attractor behaviors are noticed from the plots for $a=1.01 > 1$ (first column) and $a=0.99 < 1$ (third column), respectively.

\begin{figure}
\vspace{-1.2 cm}
\includegraphics[scale=0.32]{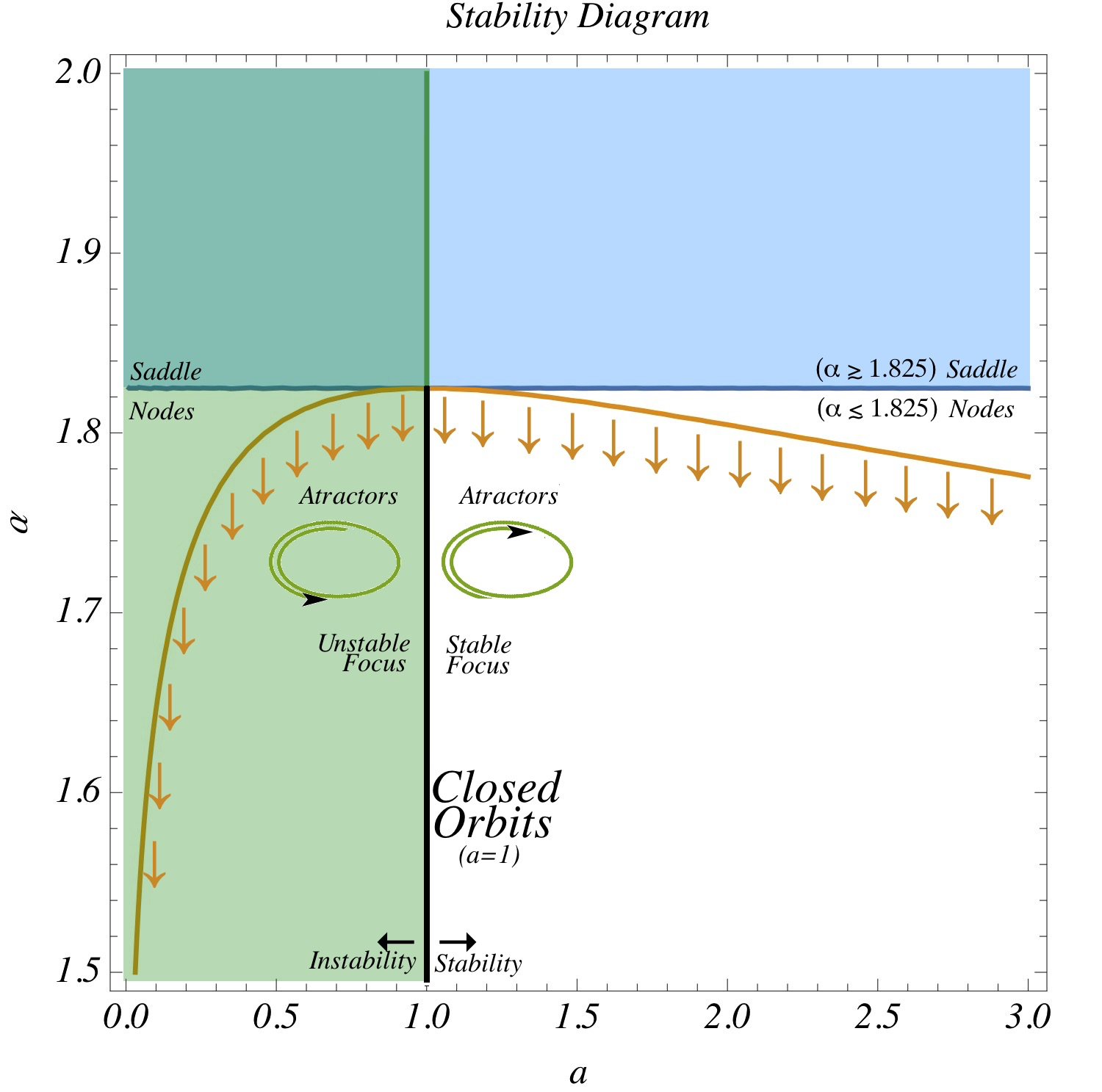}
\renewcommand{\baselinestretch}{.6}
\caption{\footnotesize
(Color online)
Extrapolated hyperbolic pattern for effective quantum prey-predator system from Eqs.~(\ref{imWA4CCD4mm}) and (\ref{imWB4CCD4mm}) as function of the $\alpha$ and the anisotropy parameter $a$.
\label{Bio0004}
}
\end{figure}

The complete hyperbolic equilibrium pattern is summarized by the results from Fig.~\ref{Bio0004}: phase-space saddle points are identified by $\alpha \gtrsim 1.825$, obtained from $Det[j(x_e,\,k_e)] < 0$; focus and node points are identified by $\alpha \lesssim 1.825$, from $Det[j(x_e,\,k_e)] > 0$; and the attractor regimes, for $\Delta[j(x_e,\,k_e)]>0$, are all defined and identified in terms of the anisotropy factor, $a$, which also sets the threshold for close orbit stability at $a=1$.
For increasing values of $\alpha$, the saddle-points which emerge from the lighter white patterns indicated in Fig.~\ref{Bio0002} naturally contributes to the subsequent diffusive appearance of unstable vortices and additional saddle-points that completely annihilate the classical pattern, where the red bubble regions correspond to the quantum drivers for the pattern depicted in the first rows of Fig.~\ref{Bio01}. However, in this case, the hyperbolic equilibrium classification corresponds to an extrapolation from the perturbative regime.

Interestingly, out of the continuous hyperbolic equilibrium domain, the multifocal vortex and saddle-point patterns of the quantum regime depicted by Fig.~\ref{Bio0002} have a chaotic counterpart in the discrete domain. 
If discreteness is admitted for the driver of prey-predator oscillation evolution, turning the LV system into a iterative discrete map, such chaotic features emerge from a single bifurcation pattern of the classical background, which is chaotically affected by quantum fluctuations, as it shall be discussed in the next section.

\section{Discretized systems -- Bifurcation maps and chaotic features}

Deterministic chaos is associated to non-linearity in Hamiltonian systems and its pattern emerges when a dynamical system depends, in a sensitive way, on its previous state.
Circumstantially, it is manifested by the discretization of the time evolving coordinate.
The minimum complexity of a chaotic system can be described by the bifurcation theory, through which changes onto the topological structure of a dynamical system can be addressed \cite{HG,Book,Book2}.
That is the case of the phase-space prey-predator-like dynamical system here investigated, when one considers that time evolves through discretized steps.

\begin{figure}[h!]
$a)$\includegraphics[scale=0.6]{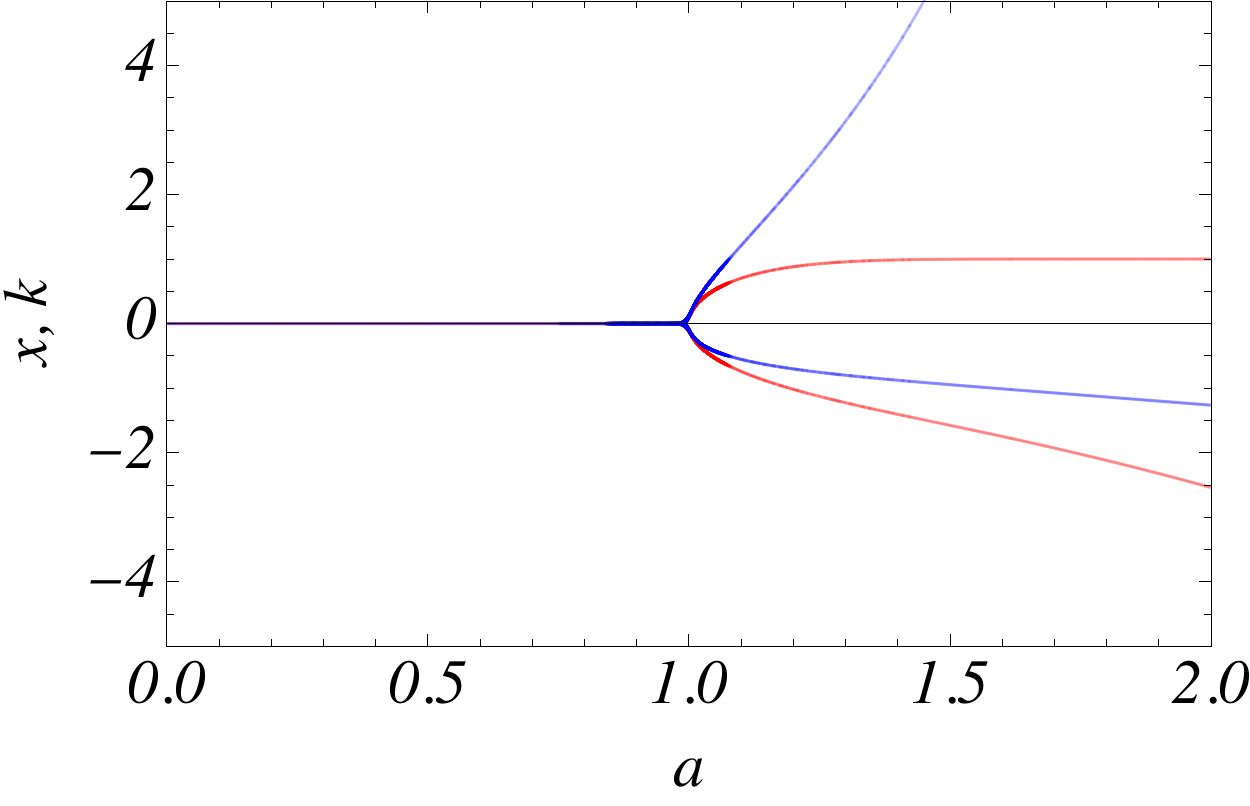}
$b)$\includegraphics[scale=0.6]{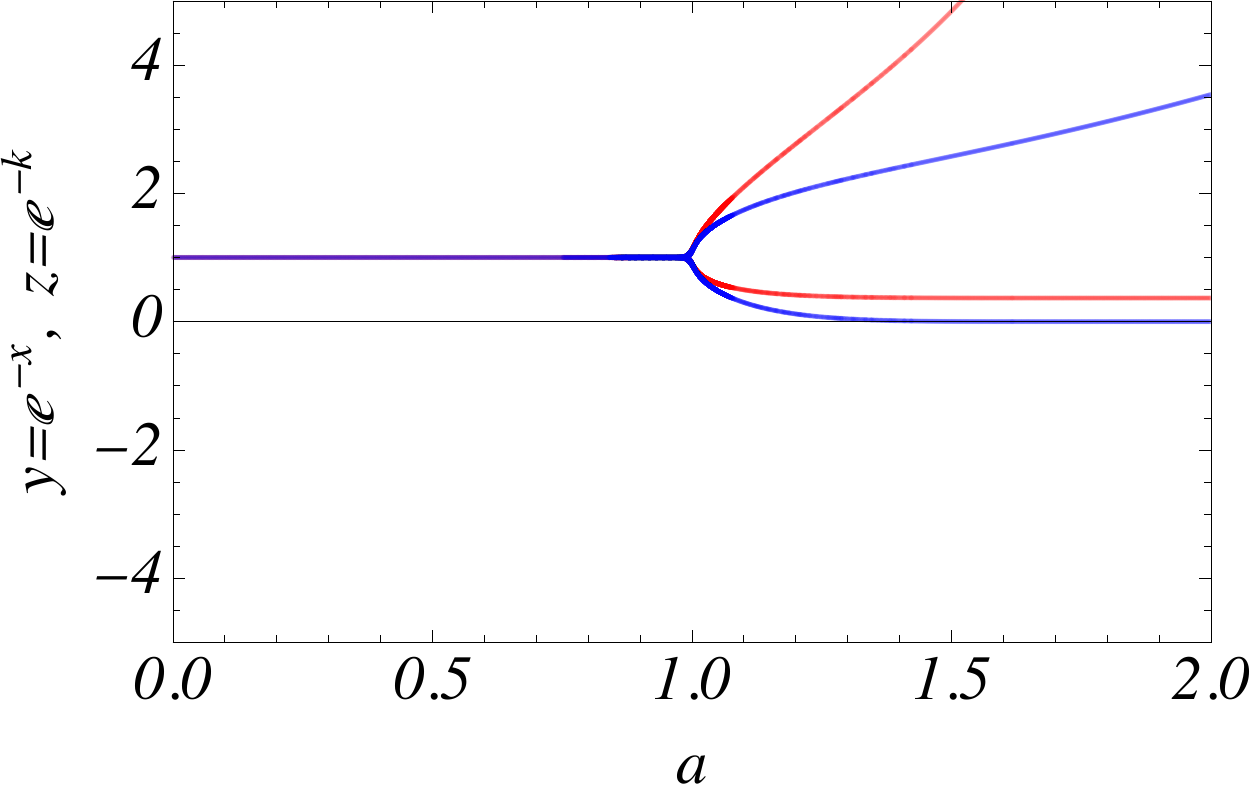}
\renewcommand{\baselinestretch}{.6}
\caption{\footnotesize
(Color online)
$a)$ (Left) Bifurcation pattern for the discrete classical evolution of the mapped prey-predator chain, $x$ and $k$, as function of the anisotropy localization parameter $a$. Red points are for the evolution of the equilibrium points for the $x$ coordinate, and blue points are for the evolution of the equilibrium points for the $k$ coordinate. Results are for $t \mapsto n/\Delta$ with $\Delta = 5000$, and $n = 1,\,2,\,\dots,\, 5\times 10^5$. $b)$ (Right) Once mapped into the prey-predator scheme,$z \mapsto e^{-k}$ and $y \mapsto e^{-x}$, the discretized pattern introduces the possibility of the prey(predator) extinction hypothesis, as the blue curve approaches zero for increasing values of $a (> 1)$.}
\label{ChaosH}
\end{figure}

According to the established theoretical grounds \cite{HG,Book,Book2}, a local bifurcation occurs when a parameter change causes the stability of an equilibrium (or fixed point) to change. In continuous systems, it is driven by the real part of an eigenvalue of an equilibrium configuration passing through zero. It results into the bifurcation point, for which the equilibrium is non-hyperbolic.
In a discrete version of both the classical and the quantum modified Gaussian convoluted LV equations, one can read $\{x(t),\, k(t)\}\mapsto\{x_n,\, k_n\}$ for $t \mapsto n/\Delta$, with $\Delta$ arbitrarily large.
Considering the results obtained in the previous section, and the natural deviations from the hyperbolic domain,
Fig.~\ref{ChaosH} depicts the bifurcation map for the prey-predator classical pattern as function of the anisotropy parameter $a$.

\begin{figure}
\vspace{-.5cm} 
\includegraphics[scale=0.07]{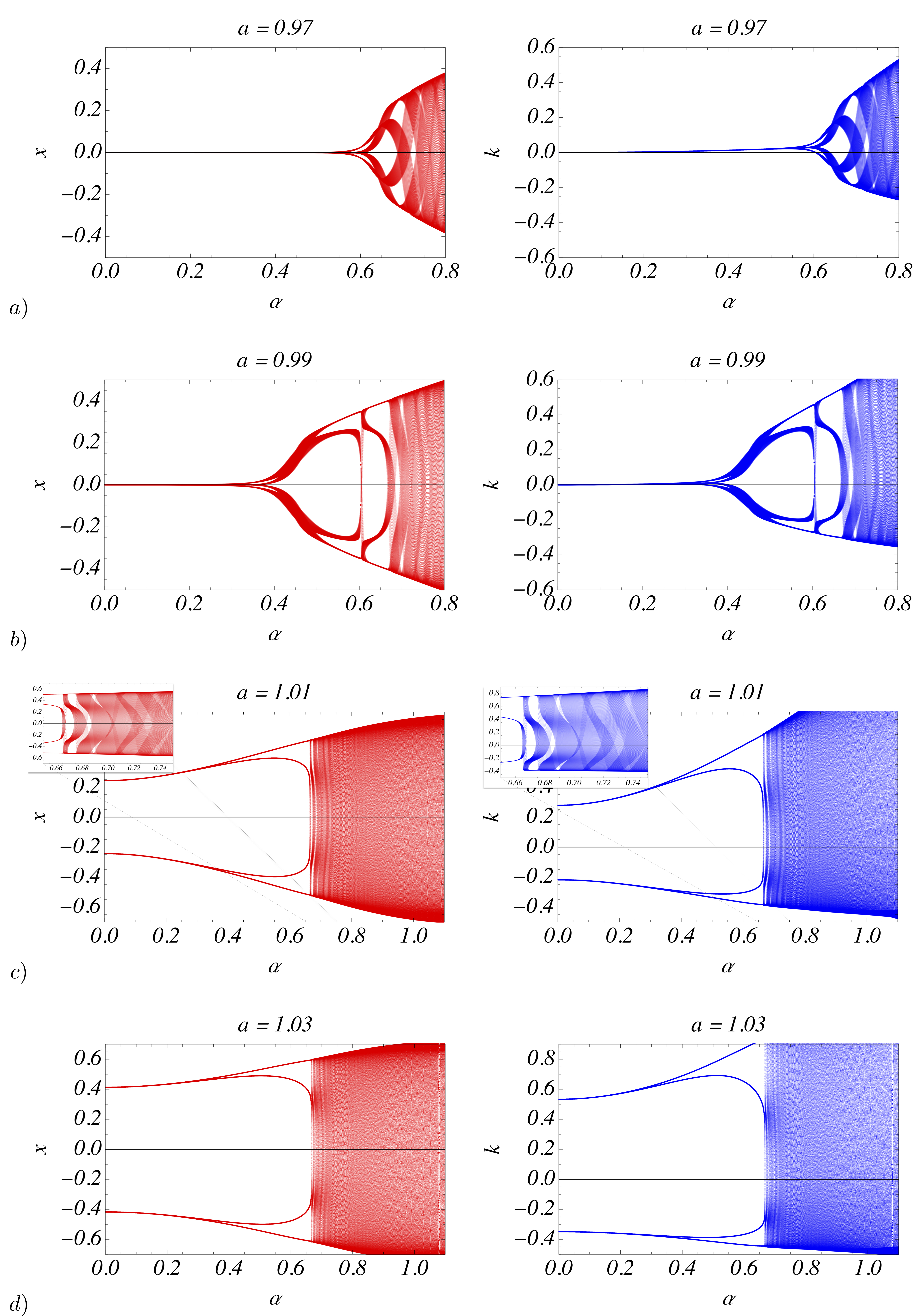}
\renewcommand{\baselinestretch}{.6}
\caption{\footnotesize
(Color online)
Bifurcation maps for: a) $a = 0.97$; b) $0.99$; c) $1.01$; and d) $1.03$; as function of the Gaussian localization parameter $\alpha$.
\label{Bio0005}. Red points are for the evolution of the equilibrium points for the $x$ coordinate, and blue points are for the evolution of the equilibrium points for the $k$ coordinate. Increasing values of $\alpha$ contribute to a complete disorder behavior which destroys classical prey-predator bifurcation patterns. Results are for $\Delta t \mapsto \epsilon$ and $t \mapsto \epsilon n$ with $\epsilon = 5 \times 10^{-4}$, and $n = 0,\,1,\,2,\,\dots$.}
\label{Chaos}
\end{figure}

One notices that the bifurcation emerge for $a = 1$ such that for $a > 1$ a two-fold degeneracy of the equilibrium point can be identified. 
Accordingly, under the influence of quantum fluctuations driven by the Gaussian parameter $\alpha$, the bifurcation at $a = 1$ can be correlated with the results for the Poincar\'e map from Fig.~\ref{Poinc}, as depicted by Fig.~\ref{Chaos}.
For increasing values of $\alpha$, the behavior of the stability points for the anisotropy parameter $a = 0.99\,(< 1)$ and $a =1.01\, (> 1)$ exhibit a two-dimensional chaotic pattern which can be indistinctly extended for any value of $a > 1$ and $a < 1$, where disorder increases with increasing values of $\alpha$.
Conversely, for $\alpha\lesssim 1$, in the first and second rows of Fig.~\ref{Chaos} ($a \lesssim 1$), just a tiny deviation from the classical equilibrium point, i.e., $(x_e,\,k_e) \sim (0,0)$, is noticed. Likewise, in the third and forth rows of Fig.~\ref{Chaos} ($a \gtrsim 1$), the classical bifurcation pattern is slightly affected by quantum corrections at $\alpha \lesssim 1$.

The topological changes in the phase-space portrait of the system define the classification of bifurcations, which by itself is extensively analyzed in the literature \cite{HG,Book,Book2} and is out of the scope of such a preliminary analysis.
From the disorder pattern numerically obtained, it can be noticed that increasing values of $\alpha$ contribute to ``accelerate'' the bifurcation pattern with respect to the anisotropy parameter, $a$, where the relative $a$-{\em distance} between successive bifurcations decreases as it is favored by the introduction of quantum distortions. 
From such results, one can realize that, for the prey-predator-like dynamics, the chaotic behavior is increasingly induced by quantum distortions.

\section{Conclusions} 

The Wigner flow framework for addressing the role of quantum fluctuations on the hyperbolic equilibrium configurations of prey-predator-like systems, as well as for understanding their relations with the triggers for chaotic patterns, has been evaluated.
Firstly, in the broad range context covered by the Wigner framework for non-linear Hamiltonians like $\mathcal{H} = \mathcal{K}(k)+\mathcal{V}(x)$, a consistent map of hyperbolic stability conditions for prey-predator-like systems was provided.
Throughout the results here obtained, quantum fluctuations over the prey-predator-like classical background has been shown to affect the pattern of hyperbolic equilibrium of their quantum analogs.
Conversely, the combination of the hyperbolic stability quantifiers with the Wigner features is shown to be relevant in distinguishing quantum fluctuations from non-linear effects when they cannot be systematically investigated through the Schr\"odinger framework.

From a theoretical perspective, the results obtained in Sec.~III could be read as the macroscopic emergence of quantum distortions over the classical pattern. The {\em quantum analog} non-commutative property related to phase-space coordinates, $[x,\,k]\neq 0$, is reflected by the non-extinction hypothesis parameterized by minimal phase-space elementary unity, $\delta x \, \delta k \sim 1$.
As a matter of fact, in quantum mechanics, position and momentum operators have their non-commutative nature expressed by the Moyal {\em star}-product definition which, by the way, recovers the WW formalism.
If $x$ and $k$ measurements affect each other, the {\em quantum analog} of the uncertainty principle is expressed by $\delta x\,\delta k \gtrsim 1$ in correspondence with $[x,\,k] = i$.
Paradigmatically, if the LV maps the prey-predator species oscillation, it is realized as a suppression of any deterministic species evolution parameterized either by $y(\tau)$ and $z(\tau)$ or by $y(z)\leftrightarrow z(y)$. However, the Wigner framework circumvents the deterministic solution by exhibiting a semiclassical (averaged-out) solution from which the quantum imprints can be detected via the hyperbolic equilibrium pattern as well as onto Poincar\'e maps and bifurcation maps (for time discretization).
From a non-deterministic perspective, competitive species, in analogy with interacting quantum states \cite{Bio17}, have their existence predictability -- which should be related to a measurement operation -- associated to a quantum statistical ensemble description. In fact, that is the case of the prey-predator-like dynamics here considered, where a Gaussian (single- or multi-particle) phase-space probability distribution is considered.
Such an analysis is consistent with the framework where several competitive microbiological and biochemical systems \cite{0001,0002} face environmental effects \cite{0003} and complex and self-organizing hierarchical mechanisms \cite{0004} included in their dynamics by means of quantum-based treatments supported by fundamental single- and multi-particle quantum mechanics.

Finally, when the above analysis is extrapolated to non-hyperbolic bifurcation regimes, the values probed or approached asymptotically by prey-predator anisotropy and Gaussian localization parameters, $a$ and $\alpha$, allowed for the complete visualization of the bifurcation theory and the correspondence between classical and effective quantum regimes.

To conclude, our results show that the LV system, as a particular example of a broad class of non-linear Hamiltonian systems, once admitted as a map of prey-predator dynamics sensitive to quantum mechanical corrections and equilibrium theory analysis, exhibit measurable patterns which, in the context of the WW phase-space framework, can be detected either microscopically by the identification of quantum topological phases or macroscopically by time evolved averaged-out statistical imprints, aspects that, of course, may deserve a fine-tuning phenomenological analysis.

\vspace{.5 cm}
{\em Acknowledgments -- The work of AEB is supported by the Brazilian Agencies FAPESP (Grant No. 20/01976-5 and Grant No. 23/00392-8) and CNPq (Grant No. 301485/2022-4).}

\section*{Appendix I -- The WW {\em quasi}-probability distribution}

The WW formalism is thought of as the bridge between operator methods and path integral techniques \cite{Abr65,Sch81,Par88} encoded by a Weyl transform operation defined by
\begin{equation}
O^W(q,\, p)\label{Wem2311}
= 2\hspace{-.15cm} \int^{+\infty}_{-\infty} \hspace{-.35cm}ds\,\exp{\left[2\,i \,p\, s/\hbar\right]}\,\langle q - s | \hat{O} | q + s \rangle=2\hspace{-.15cm} \int^{+\infty}_{-\infty} \hspace{-.35cm} dr \,\exp{\left[-2\, i \,q\, r/\hbar\right]}\,\langle p - r | \hat{O} | p + r\rangle,
\end{equation} 
where an arbitrary quantum operator is identified by $\hat{O}$.
For the case where $\hat{O}$ is identified as a density matrix operator, $\hat{\rho} = |\psi \rangle \langle \psi |$, the Weyl transformed operator, $O^W(q,\, p)$, results in the so-called Wigner {\em quasi}-probability distribution function,
\begin{equation}
 h^{-1} \hat{\rho} \to W(q,\, p) = (\pi\hbar)^{-1} 
\int^{+\infty}_{-\infty} \hspace{-.35cm}ds\,\exp{\left[2\, i \, p \,s/\hbar\right]}\,
\psi(q - s)\,\psi^{\ast}(q + s).\label{Wem23}
\end{equation}
This is interpreted as the Fourier transform of the off-diagonal elements of the associated density matrix, $\hat{\rho}$, where $h = 2\pi \hbar$ is the Planck constant. As a mandatory constraint, it presumes a consistent probability distribution interpretation constrained by the normalization condition over $\hat{\rho}$, that is $Tr_{\{q,p\}}[\hat{\rho}]=1$.

Eq.~\eqref{Wem23} was proposed by Wigner's seminal work \cite{Wigner} when accounting for quantum corrections to TD equilibrium states.
Without affecting the predictive power of QM, and preserving the symmetries of the Heisenberg-Weyl group of translations, the Wigner phase-space {\em quasi}-distribution function is therefore associated with a density operator $\hat{\rho}$, in the form of an overlap integral, Eq.~\eqref{Wem23}.
The reason for the {\em quasi}-distribution nomenclature is due to the point that the Wigner function's most elementary property is concerned with its marginal distributions, which return position and momentum distributions upon integrations over the momentum and position coordinates, respectively,
\begin{equation}
\vert \psi(q)\vert^2 = \int^{+\infty}_{-\infty} \hspace{-.35cm}dp\,W(q,\, p)
\qquad
\leftrightarrow
\qquad
\vert \varphi(p)\vert^2 = \int^{+\infty}_{-\infty} \hspace{-.35cm}dq\,W(q,\, p).
\end{equation}
In fact, strictly connected with the Hilbert space features of the Schr\"odinger QM, the Fourier transform of the above addressed wave functions,
\begin{equation}
 \varphi(p)=
(2\pi\hbar)^{-1/2}\int^{+\infty}_{-\infty} \hspace{-.35cm} dq\,\exp{\left[i \, p \,q/\hbar\right]}\,
\psi(q),
\end{equation}
is the property that suppresses the coexistence of positive-definite position and/or momentum probability distributions.

Hence, the connection of the Wigner function to the matrix operator QM, through Eqs.~\eqref{Wem2311} and \eqref{Wem23}), allows for computing the expected values of quantum observables described by generic operators, $\hat{A}$, evaluated through an overlap integral over the infinite volume described by the phase-space coordinates, $q$ and $p$, as \cite{Wigner,Case}
\begin{equation}
 \langle O \rangle = 
\int^{+\infty}_{-\infty} \hspace{-.35cm}dp\int^{+\infty}_{-\infty} \hspace{-.35cm} {dq}\,\,W(q,\, p)\,{A^W}(q,\, p). \label{MOREfive}
\end{equation}
This corresponds to the trace of the product between $\hat{\rho}$ and $\hat{O}$, $Tr_{\{q,p\}}\left[\hat{\rho}\hat{O}\right]$.
In addition, the statistical aspects evinced from the definition of $W(q,\, p)$ admit extensions from pure states to statistical mixtures, through which the replacement of ${O^W}(q,\, p)$ by $W(q,\, p)$ into Eq.~\eqref{MOREfive} leads to the quantum purity computed through an analogous of the trace operation, $Tr_{\{q,p\}}[\hat{\rho}^2]$, i.e.
\begin{equation}
Tr_{\{q,p\}}[\hat{\rho}^2] = 2\pi\hbar\int^{+\infty}_{-\infty}\hspace{-.35cm}dp\int^{+\infty}_{-\infty} \hspace{-.35cm} {dq}\,\,\,W(q,\, p)^2,
\label{MOREpureza}
\end{equation}
which satisfies the pure state constraint, $Tr_{\{q,p\}}[\hat{\rho}^2] = Tr_{\{q,p\}}[\hat{\rho}] = 1$.

Finally, flow properties of the Wigner function, $W(q,\,p) \to W(q,\,p;\,t)$, can be, for instance, connected to the Hamiltonian dynamics.
In this case, a vector flux \cite{Steuernagel3,NossoPaper,Meu2018}, $\mathbf{J}(q,\,p;\,t)$, decomposed into the phase-space coordinate directions, $\hat{q}$ and $\hat{p}$, as $\mathbf{J} = J_q\,\hat{q} + J_p\,\hat{p}$,
reproduces a flow field connected to the Wigner function dynamics through the continuity equation \cite{Case,Ballentine,Steuernagel3,NossoPaper,Meu2018},
\begin{equation}
{\partial_t W} + {\partial_q J_q}+{\partial_p J_p} =0,
\label{alexquaz51}
\end{equation}
where a shortened notation for partial derivatives, $\partial_a \equiv \partial/\partial a$, has been used.

To connect the framework with prey-predator dynamics, all the above quantities have been more properly written in terms of dimensionless variables, $x = \left(m\,\omega\,\hbar^{-1}\right)^{1/2} q$ and $k = \left(m\,\omega\,\hbar\right)^{-1/2}p$. In this case, one should have the dimensionless Hamiltonian $\mathcal{H} = (\hbar \omega)^{-1} H (q,\,p) = \mathcal{K}(k)+\mathcal{V}(x)$, with $\mathcal{V}(x) = (\hbar \omega)^{-1} V\left(q\right)=(\hbar \omega)^{-1} V\left(\left(m\,\omega\,\hbar^{-1}\right)^{-1/2}x\right)$ and $\mathcal{K}(k) = (\hbar \omega)^{-1} K\left(p\right)= (\hbar \omega)^{-1} K\left(\left(m\,\omega\,\hbar\right)^{1/2}k\right)$, where $m$ is a mass scale parameter and $\omega$ is an arbitrary angular frequency.
Therefore the Wigner function can be cast into the dimensionless form of $\mathcal{W}(x, \, k;\,\omega t) \equiv \hbar\, W(q,\,p;\,t)$, with $\hbar$ absorbed by $dp\,dq\to \hbar\, dx\,dk$ integrations,
\begin{eqnarray}\label{alexDimW} 
\mathcal{W}(x, \, k;\,\tau) &=& \pi^{-1} \int^{+\infty}_{-\infty} \hspace{-.35cm}dw\,\exp{\left(2\, i \, k \,w\right)}\,\psi(x - w;\,\tau)\,\psi^{\ast}(x + w;\,\tau),
\end{eqnarray}
with $w = \left(m\,\omega\,\hbar^{-1}\right)^{1/2} s$ and $\tau = \omega t$, as it appears in Eq.~\eqref{222} with $\tau$ suppressed from notation.

\end{document}